\pdfoutput = 1

\documentclass[aps,pre,twocolumn,superscriptaddress,showpacs]{revtex4}

\usepackage{amsfonts,amssymb,amsmath,latexsym,epsfig,wasysym}
\usepackage[sort&compress]{natbib}
\usepackage{bm}
\usepackage{hyperref}
\usepackage[latin1]{inputenc}
\usepackage{ulem}
\usepackage{multirow}
\usepackage{rotating}

\usepackage{color}
\definecolor{greencolor}{rgb}{0,0.5,0.2}
\definecolor{redcolor}{rgb}{.7,0.,0.}
\definecolor{bluecolor}{rgb}{0,0.,1.}
\definecolor{greycolor}{rgb}{.5,.5,.5}

\begin{document}

\title{Probing the statistical properties of unknown texts:
application to the Voynich Manuscript}

\author{Diego R. Amancio}
\affiliation{
Institute of Physics of S\~ao Carlos,
University of S\~ao Paulo,
P. O. Box 369,
Postal Code 13560-970,
S\~ao Carlos,
S\~ao Paulo,
Brazil}

\author{Eduardo G. Altmann}
\affiliation{Max Planck Institute for the Physics of Complex Systems (MPIPKS),
01187 Dresden,
Germany}

\author{Diego Rybski}
\affiliation{
Potsdam Institute for Climate Impact Research (PIK),
P.O. Box 60 12 03,
14412 Potsdam,
Germany}

\author{Osvaldo N. Oliveira Jr.}
\affiliation{
Institute of Physics of S\~ao Carlos,
University of S\~ao Paulo,
P. O. Box 369,
Postal Code 13560-970,
S\~ao Carlos,
S\~ao Paulo,
Brazil}

\author{Luciano da F. Costa}
\affiliation{
Institute of Physics of S\~ao Carlos,
University of S\~ao Paulo,
P. O. Box 369,
Postal Code 13560-970,
S\~ao Carlos,
S\~ao Paulo,
Brazil}


\begin{abstract}
While the use of statistical physics methods to analyze large corpora has been useful to unveil many patterns in texts, no comprehensive investigation has been performed investigating the properties of statistical measurements
across different languages and texts. In this study we propose a framework that aims at determining
if a text is compatible with a natural language and which languages are closest
to it, without any knowledge of the meaning of the words. The approach is based
on three types of statistical measurements, i.e.\ obtained from first-order
statistics of word properties in a text, from the topology of complex networks
representing text, and from intermittency concepts where text is treated as a
time series. Comparative experiments were performed with the New Testament in 15
different languages and with distinct books in English and Portuguese in order
to quantify the dependency of the different measurements on the language and on
the story being told in the book. The metrics found to be informative in
distinguishing real texts from their shuffled versions include assortativity,
degree and selectivity of words. As an illustration, we analyze an undeciphered
medieval manuscript known as the Voynich Manuscript. We show that it is mostly
compatible with natural languages and incompatible with random texts. We also
obtain candidates for key-words of the Voynich Manuscript which could be helpful
in the effort of deciphering it. Because we were able to identify statistical
measurements that are more dependent on the syntax than on the semantics, the
framework may also serve for text analysis in language-dependent applications.
\end{abstract}


\maketitle


\tableofcontents

\section{Introduction}\label{sec.intro}

Methods from statistics, statistical physics, and artificial intelligence have
increasingly been used to analyze large volumes of text for a variety of
applications~\cite{twitter,Michel,rede3,comparing,kkk,patterns,smallWorld}
some of which are related to fundamental linguistic and cultural phenomena.
Examples of studies on human behaviour are the analysis of
mood change in social networks~\cite{twitter}
and the identification of literary
movements~\cite{rede3}. Other applications of statistical natural language processing techniques include the development of statistical techniques to improve the performance of information retrieval systems~\cite{ir}, search engines~\cite{searchEngines}, machine
translators~\cite{mt,cnanalysis} and automatic summarizers~\cite{suma}.
Evidence of the success of statistical techniques for natural language
processing is the superiority of current corpus-based machine translation
systems in comparison to their counterparts based on the symbolic approach~\cite{simbolico}.

The methods for text analysis we consider can be classified into three broad classes:
(i) those based on first-order statistics where data on classes of words are used in the
analysis, e.g.\ frequency of words~\cite{manning};
(ii) those based on metrics from
networks representing
text~\cite{smallWorld,patterns,rede3,rede4,comparing};
(iii) those using intermittency concepts and time-series analysis for
texts~\cite{comparing,kkk}. One of the major advantages inherent in these
methods is that no knowledge about the meaning of the words or the syntax of the
languages is required. Furthermore, large corpora can be processed at once, thus
allowing one to unveil hidden text properties that would not be probed in a
manual analysis given the limited processing capacity of humans. The obvious
disadvantages are related to the superficial nature of the analysis, for even
simple linguistic phenomena such as lexical disambiguation of homonymous words
are very hard to treat. Another limitation in these statistical methods is the
need to identify the representative features for the phenomena under
investigation, since many parameters can be extracted from the analysis but
there is no rule to determine which are really informative for the task at hand.
Most significantly, in a statistical analysis one may not even be sure if the
sequence of words in the dataset represents a meaningful text at all. For testing whether an unknown text is compatible with natural language,
one may calculate measurements for this text and several others of a known language, and then verify if the results are statistically compatible. However, there may be variability among texts of the same language, especially owing to semantic issues.


In this study we combine measurements from the three classes above and propose a framework to determine the importance of these measurements in investigations of unknown texts, regardless of the alphabet in which the text is encoded. The statistical properties of words and the books were obtained for comparative studies involving the same book (New Testament) in 15 languages and distinct pieces of text written in English and Portuguese. The purpose in this type
of comparison was to identify the features capable of distinguishing a
meaningful text from its shuffled version (where the position of the words is randomized), and then determine the proximity of pieces of text.

As an application of the framework, we analyzed the famous
Voynich Manuscript (VMS), which has remained indecipherable in spite of attempts from renowned cryptographers for a century. This manuscript dates back to the 15th century,
possibly produced in Italy, and was named after Wilfrid Voynich who bought it in
1912. In the analysis we make no attempt to decipher VMS, but
we have been able to verify that it is compatible with natural languages, and
even identified important keywords, which may provide a useful starting point
toward deciphering it.

\section{Description of the measurements}\label{definicao}

\begin{figure*}[bt]
  \centering
  \includegraphics[width=1.8\columnwidth]{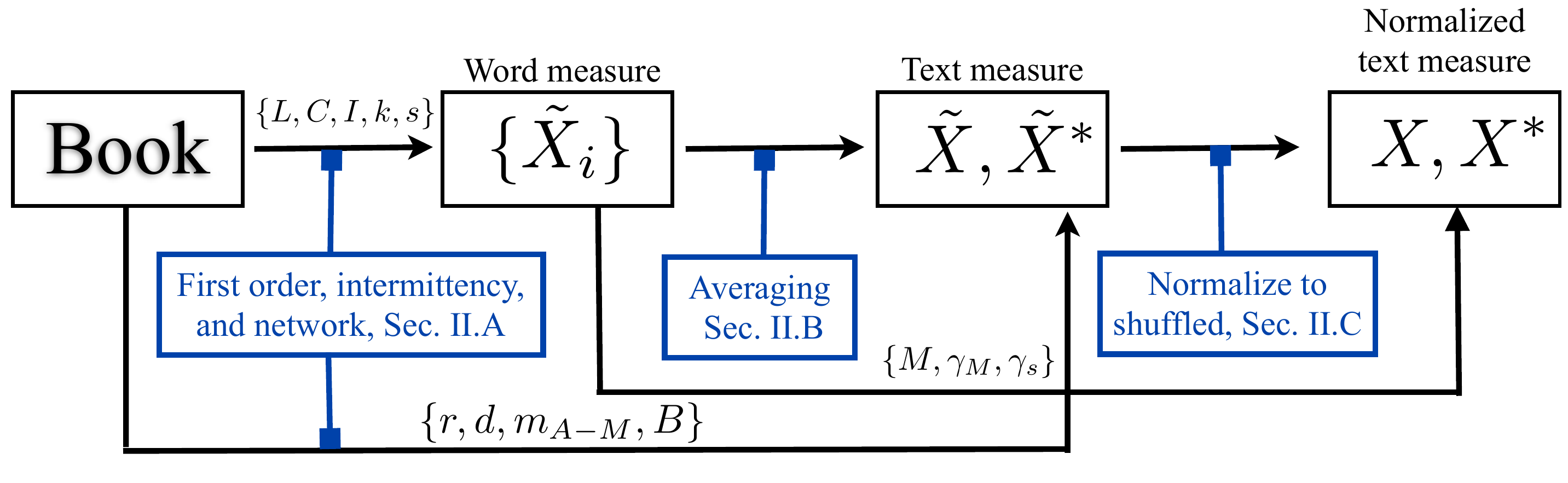}
  \caption{Illustration of the procedures performed to obtain a measurement $X$ of each book. \label{fig.ilust}}
\end{figure*}

The analysis involves a set of steps going beyond the
basic calculation of measurements, as illustrated in the workflow in Fig.~\ref{fig.ilust}. Some measurements are averaged in order to obtain
a measurement on the text level from the measurement on the word level. In addition, a comparison with values obtained after randomly shuffling the text is performed to assess to which extent structure is reflected in the measurements.

\subsection{Raw measurements}\label{ssec:rawmeasures}

\subsubsection{First order statistics}

The simplest measurements obtained are the vocabulary size $M$, which is the number of distinct words in the text, and the frequency of word~$i$ (number of appearances), denoted by $N_i$. The heterogeneity of the contexts surrounding words was quantified with the so-called selectivity measurement~\cite{selet}. If a word is strongly selective then it always co-occurs with the same adjacent words. Mathematically, the selectivity of a word $i$ is $s_i = 2 N_i/t_i$, where $t_i$ is the number of distinct words that appear immediately beside (i.e., before or after) $i$ in the text.

A language-dependent feature is the number of different words (types) that at least once had two word tokens immediately beside each other in the text. In some languages this repetition is rather unusual, but
in others it may occur with a reasonable frequency (see Sec.~\ref{resultados} and Figure \ref{img:example}). In this paper, the number of repeated bigrams is denoted by $B$.

\subsubsection{Network characterization}

Complex networks have been used to characterize texts~\cite{smallWorld,patterns,rede3,rede4,comparing},
where the nodes represent words and links are established based on word co-occurrence,
i.e.\ links between two nodes are established if the corresponding words appear at least once adjacent in the text.
$j$.
%
%
%
In most applications of co-occurrence networks, the
stopwords~\footnote{Stopwords are highly frequent words usually conveying little
semantic information, e.g.\ articles and prepositions.} are removed and the
remaining words are lemmatized~\footnote{The lemmatization consists in
transforming the word to its canonical form. Thus conjugated verbs are converted
to their infinitive form and plural nouns are converted to their singular
form.}. Here, we decided not to do this because in unknown languages it is
impossible to derive lemmatized word forms or identify stopwords.
To characterize the structure and organization of the networks, the following topological
metrics of complex networks were calculated
(more details are given in the Supplementary Information (SI)):

\begin{itemize}

\item
We quantify \emph{degree correlations}, i.e.\ the tendency of nodes of certain degree to be connected to nodes with similar degree (the degree of a node is the number of links it has to other nodes), with the Pearson correlation coefficient, $r$, thus distinguishing assortative ($r>1$) from disassortative ($r<1$) networks.

\item
The so-called clustering coefficient, $C_i$, is given by the fraction of
closed triangles of a node, i.e.\ the number of actual connections between
neighbours of a node divided by the possible number of connections between
them. The global \emph{clustering coefficient}~$C$ is the average over the
local coefficients of all nodes.

\item
The \emph{average shortest path length}, {$L_i$, is the shortest path between two nodes~$i$ and~$j$ averaged over all possible $j$'s.}
In text networks it measures the relevance of words
according to their distance to the most frequent words~\cite{comparing}.

\item
The \emph{diameter} $d$ corresponds to the maximum shortest path,
i.e.\ the maximum distance on the network between any two nodes.

\item
We also characterized the topology of the networks
through the analysis of motifs, i.e.\ analysis of connectivity patterns expressed
in terms of small building blocks (or subgraphs)~\cite{milo}.
We define as $m_Y$ the number of motifs $Y$ appearing in the network.
The motifs employed in the current paper are displayed in Figure \ref{img:tourist-walk-schematic}.

\begin{figure}[h]
  \centering
  \includegraphics[scale=0.45]{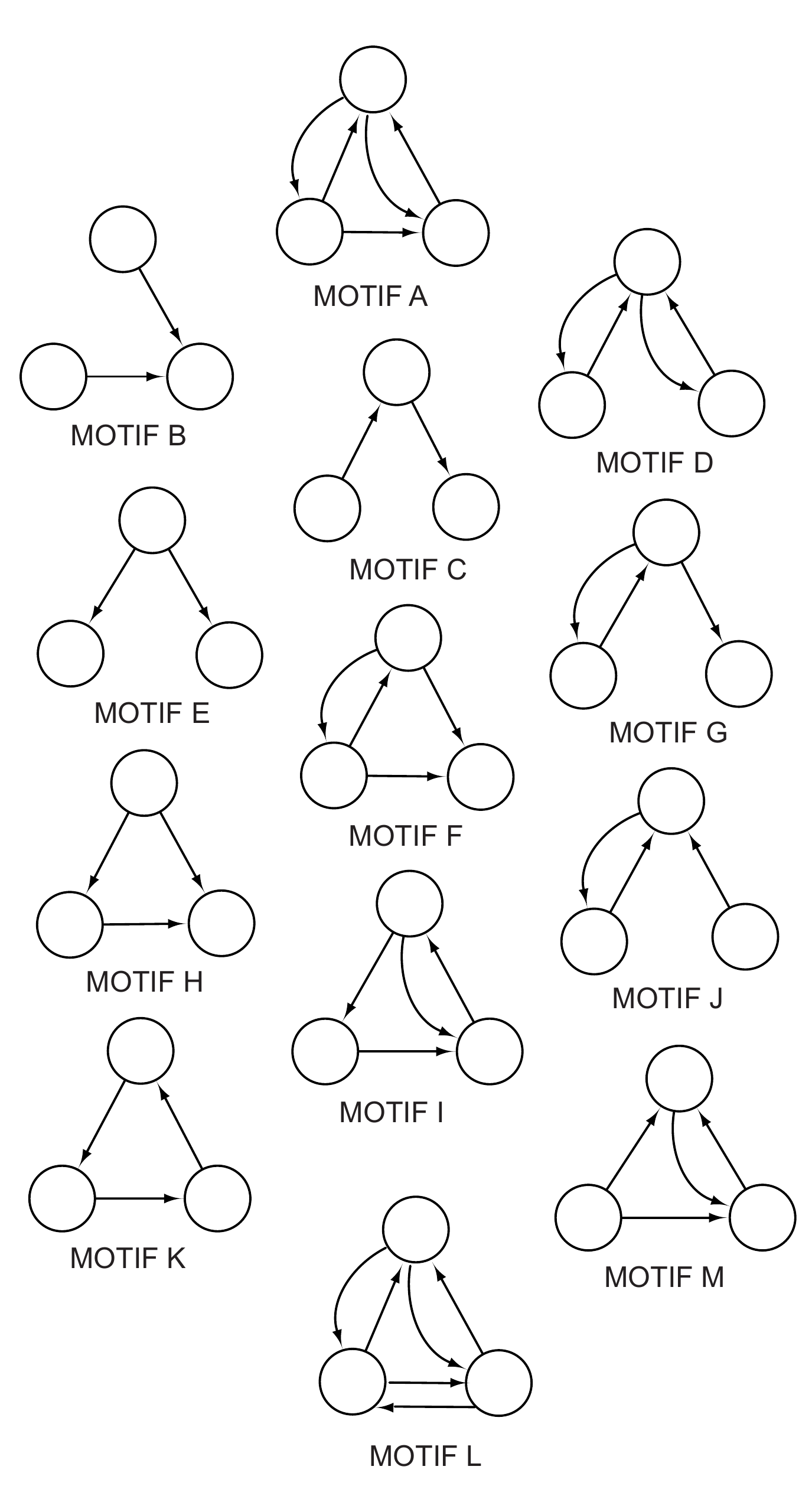}
  \caption{Illustration of $13$ motifs comprising three nodes used to analyze
  		the structure of text networks.}
  \label{img:tourist-walk-schematic}
\end{figure}



\end{itemize}

\subsubsection{Intermittency} \label{sec:intermittency}

The fact that words are unevenly distributed along texts has been used to detect
keywords in documents~\cite{entropy-keyword,key1,kkk}.
Since bursty words appear concentrated in portions of the text in contrast to others, which are distributed homogenouly along the text, words with different functions can be distinguished.

The intermittency was calculated using the concept of recurrence times, which
have been used to quantify the burstiness of time series. In the case of
documents, the time series of a word is taken by counting the number of words
(representing time) between successive appearances of the considered word. For
example, the recurrence times for the word `the' in the previous sentence are
$T_1 = 4, T_2 = 10,$ and $T_3=11$. If $N_i$ is the frequency of the word its time series
will be composed by the following elements \{$T_1$, $T_2$, $\ldots$
$T_{N_i-1}$\}. Because the times until the first occurrence $T_f$ and after the
last occurrence $T_l$ are not considered, the element $T_N$ is arbitrarily
defined as $T_N = T_f + T_l$. Note that with the inclusion of $T_N$ in the time
series, the average value over all $N_i$ values is $\langle T \rangle_i = N / N_i$. Then, to compute the heterogeneity of
the distribution of a word $i$ in the text, we obtained the intermittency $I_i$
as
\begin{equation}
    I_i = \frac{\sqrt{ \langle T^2 \rangle_i  - \langle T \rangle^2_i }}{\langle T \rangle_i}.
\end{equation}
Words distributed by chance have $I_i \simeq 1$ (for $N_i\gg1$), while bursty words have $I_i > 1$.
Words with $N_i < 5$ were neglected since they lack statistics.



\subsection{From word to text measurements}
Many of the measurements defined in the previous Section are attributes of the {\it word} $i$. For our aims here it is essential to compare different {\it texts}. The easiest and most straightforward choice is to assign to a piece of text the average value of each measurement $\tilde{X}_i$, computed over all $M$ words in the text $\tilde{X} = M^{-1} \sum \tilde{X}_i$. This was done for $L$, $C$, $I$, $k$ and $s$. One potential limitation of this approach is that the same weight is attributed to each word, regardless of their
frequency in the text. To overcome this, we also calculated another metric, $\tilde{X}^*$ obtained as the average of the
$\eta$ most frequent words, i.e.\ $\tilde{X}^* = \eta^{-1} \sum X_i$, where the sum runs over the $\eta$ most frequent words.  Here, we chose $\eta=50$. Finally, because $s$ is known to have a distribution with long tails~\cite{selet}, we also computed the coefficient $\gamma_s$ of the power-law $P(s) \propto s^{-\gamma_s}$, for which the maximum-likelihood methodology described in \cite{maximo} was used.

\subsection{Comparison to shuffled texts}

Since we are interested in measurements capable of distinguishing a meaningful text from its shuffled version, each of the measurements $\tilde{X}$ and $\tilde{X}^*$ described above was normalized by the average obtained over $10$ texts produced using a word
shuffling process, i.e.\ randomizing preserving the word frequencies. If $\mu(\tilde{X}^{(R)})$ and $\sigma(\tilde{X}^{(R)})$ are respectively the average and the deviation over $10$ realizations of shuffled texts, the normalized measurement $X$ and the uncertainty $\epsilon(X)$ related to $X$ are:
\begin{equation}
    X = \frac{\tilde{X}}{\mu(\tilde{X}^{(R)})}
\end{equation}
\begin{equation} \label{sssigma}
    \epsilon(X) = \frac{ \sigma(\tilde{X}^{(R)})}{  \mu(\tilde{X}^{(R)}) ^2} \tilde{X} = \frac{ \sigma(\tilde{X}^{(R)})}{  \mu(\tilde{X}^{(R)}) } X
\end{equation}
Normalization by the shuffled text is useful because it permits comparing each measurement with a
null model. Hence, a measurement provides significant information only if its normalized $X$ value is not $\epsilon(X)$ close to $X^* =1$. Moreover, the influence of the
vocabulary size $M$ on the other measurements tends to be minimized.

\section{Variability across languages and texts} \label{resultados}

\begin{figure}[h]
  \centering
  \includegraphics[scale=0.9]{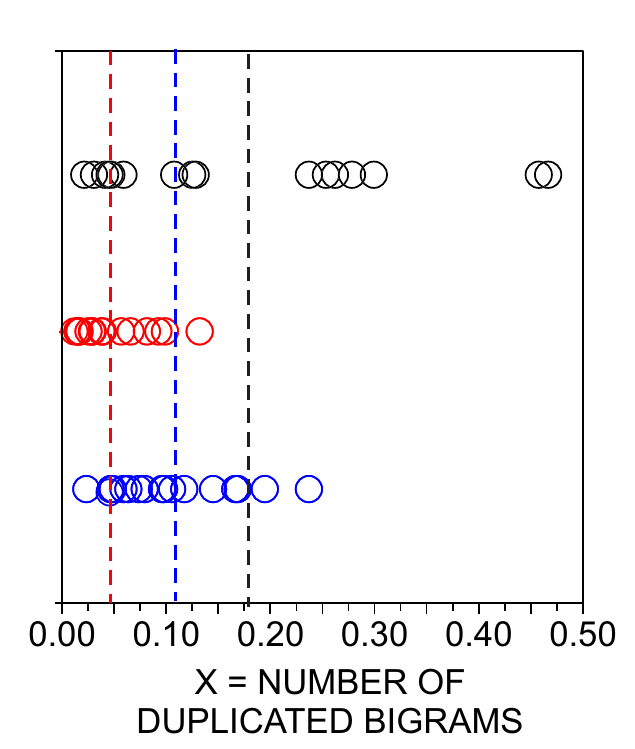}
  \caption{Distribution of $X=B$ for the New Testament (black circles), English
		(red circles) and Portuguese (blue circles) texts.
		The average $\langle X \rangle$ for the three sets of texts is represented
		as dashed lines.}
  \label{img:example}
\end{figure}

The measurements described in Section \ref{definicao} vary from text to text due
to the syntactic properties of the language. In a given language, there is also
an obvious variation among texts on account of stylistic and semantic factors.
Thus, in a first approximation one may assume that variations across texts of a measurement $X$ occur in two dimensions. Let
$X_{t,l}$ denotes the value of $X$ for text $t$ written in language $l$.
If we had access to the complete matrix $X_{t,l}$, i.e.\ if all
possible texts in every possible language could be analyzed, we could simply
compare a new text~$t$ to the full variation of the measurements $X_{t,l}$ in order, e.g., to attribute to which languages~$\lambda$
the text is compatible with. In practice, we can at best have some rows and columns filled and therefore additional statistical tests are
needed in order to characterize the variation of specific measurements.
%
For different texts, $P(X_{t,l=\lambda})$ denotes the distribution of measurement $X$ across different texts
in a fixed language $l=\lambda$ and $P(X_{t=\tau,l})$ the distribution of $X$
across a fixed text $t=\tau$ written in various languages. Accordingly, $\mu(P)$
and $\sigma(P)$ represent the expectation and the variation of the distribution
$P$. 
%
For concreteness, Fig.~\ref{img:example} illustrates the distribution of
$X=B$ (number of duplicated bigrams) for the three sets of texts we use in our analysis:  $15$ books in Portuguese, $15$ books in English,
and $15$ versions of the New Testament in different languages, see SI for details. We consider also the average $\langle X \rangle$ and the standard deviation $\sigma(X)$ of $X$ computed over different books (e.g., each of the three sets of $15$ books) and the correlation $R_M$ between $X$ and the vocabulary size $M$ of the book. Table~\ref{app:tab.1} shows the values of $\langle X \rangle, \sigma(X)$ and $R_M$ of all measurements in each of the three sets of books. In order to obtain further insights on the dependence of these measurements on language (syntax) and text (semantics), next we perform additional statistical analysis to identify measurements that are more suitable to target specific problems.

\begin{table*}
    \centering
    \caption{\label{app:tab.1}Verification of which measurements satisfy conditions $\zeta_1$, $\zeta_2$, $\zeta_2'$ and $\zeta_3$. $R_M$ is the Pearson correlation between $X$ and the vocabulary size $M$. We assume that $\zeta_1$, $\zeta_2$, $\zeta_2'$ and $\zeta_3$ are satisfied respectively when $\rho = 0.00~\%$,  $\upsilon_{t=new,l} > \upsilon_{t, l=\lambda}$, $\iota(\upsilon_{t=\tau,l}) \cap \iota(\upsilon_{t,l=\lambda}) \leq 0.05 \iota(\upsilon_{t=\tau,l}) \cup \iota(\upsilon_{t,l=\lambda})$ and $c(X_{t=new,l=\lambda},P(X_{t,l=\lambda})) > 0.05$. Measurements satisfying conditions for all three sets of texts are marked with a filled circle ($\bullet$). 
    }
    \begin{tabular}{|c|c|c|c|c|c|c|c|c|c|c|c|c|c|c|c|c|c|c|c|}
        \hline
        \multirow{2}{*}{$X$} & \multicolumn{3}{|c|}{ $\langle X \rangle \pm \sigma(X)$}  &  \multicolumn{3}{|c|}{$\rho(X=1,\{X\})$} &  \multicolumn{2}{|c|}{$\upsilon_{t=new,l} / \upsilon_{t, l=\lambda}$} & \multicolumn{2}{|c|}{$c(X,P(X))$} &\multirow{2}{*}{$R_M$} & \multirow{2}{*}{$\zeta_1$}  & \multirow{2}{*}{$\zeta_2$} & \multirow{2}{*}{$\zeta_2'$ } & \multirow{2}{*}{$\zeta_3$ } \\
        \cline{2-11}
            & $\tau = $ new &  $\lambda = $ en & $\lambda =$ pt & $\tau = $ new &  $\lambda =$ en & $\lambda = $ pt  & $\lambda = $  en & $\lambda = $  pt & $\lambda = $  en & $\lambda = $  pt &  & & &  &\\
        \hline
        $M$              & $5,809 \pm 2,665$ & $4,720 \pm 922$ & $6,921 \pm 1,126$ & -- & -- & -- &  3.12 & 2.82 & 0.00 & 0.00 & +1.00 & -- &  $\bullet$ & $\bullet$ & \\
        $\gamma_M$          & $1.99 \pm 0.11 $ & $1.93 \pm 0.06$ & $2.01 \pm 0.09$ & -- & -- & -- &  1.71 & 1.25 & 0.00 & 0.00 & +0.86 &  -- &  $\bullet$ & & \\
        $r$              &  $0.91 \pm 0.10$  & $1.10 \pm 0.06$ & $1.15 \pm 0.04 $ & $0.00~\%$ & $0.00~\%$ & $0.00~\%$ & 2.18 & 3.41 & 0.07 & 0.14 & +0.07  & $\bullet$ &  $\bullet$ & $\bullet$ & $\bullet$\\
        $d$ & $1.44 \pm 0.58$ & $1.32 \pm 0.38$ & $1.07 \pm 0.14$ & $12.50~\%$ & $37.50~\%$ & $43.75~\%$  &  1.41 & 3.16 & 0.00 & 0.00 & +0.08 &  & $\bullet$ & & \\
        $L$ &  $1.04 \pm 0.05$ & $0.99 \pm 0.02$ & $0.97 \pm 0.01 $ & $12.50~\%$ & $0.00~\%$ & $0.00~\%$ &  2.07 & 7.57 & 0.76 & 0.68 & +0.20 & & $\bullet$ & $\bullet$ & $\bullet$ \\
        $L^*$ & $1.08 \pm 0.04$ &  $1.04 \pm 0.02$ & $1.03 \pm 0.01$ & $0.00~\%$ & $0.00~\%$ & $0.00~\%$	 & 2.23 & 2.91 & 0.80 & 0.51 & +0.34 & $\bullet$ &  $\bullet$  & $\bullet$ & $\bullet$\\	
	    $ C $     &  $0.83 \pm 0.13$ & $0.97 \pm 0.04$ & $0.97 \pm 0.03$ & $0.00~\%$ & $18.75~\%$ & $25.00~\%$    & 3.31 & 4.74 & 0.65 & 0.62 & -0.34 &  &  $\bullet$ & $\bullet$ & $\bullet$ \\
        $ C^*$   &  $0.66 \pm 0.13$  & $0.65 \pm 0.08$ & $0.63 \pm 0.07$ & $0.00~\%$ & $0.00~\%$ & $0.00~\%$	 & 1.52 & 1.71 & 0.91 & 0.80 & -0.58 & $\bullet$  & $\bullet$ & & $\bullet$ \\
        $ I $     &  $1.30 \pm 0.07$ & $1.29 \pm 0.14$ & $1.27 \pm 0.06$ & $0.00~\%$ & $0.00~\%$ & $0.00~\%$	 &   0.47 & 1.03 & 0.59 & 0.45 & -0.43 & $\bullet$ & & & $\bullet$ \\
        $ I^*$   &  $1.32 \pm 0.05$ & $1.32 \pm 0.14$ & $1.26 \pm 0.09$ & $0.00~\%$ & $0.00~\%$ & $0.00~\%$    &   0.36 & 0.75 & 0.77 & 0.95 & -0.26 & $\bullet$  &  & $\bullet$ & $\bullet$ \\
        $B$                     &  $0.18 \pm 0.15$ & $0.05 \pm 0.04$ & $0.10 \pm 0.05$ & $0.00~\%$ & $0.00~\%$ & $0.00~\%$ &   1.01 & 11.4 & 0.95 & 0.32 & +0.27 & $\bullet$  & $\bullet$ & & $\bullet$ \\	
        $ k $     &  $0.71 \pm 0.06$ & $0.82 \pm 0.03$ & $0.87 \pm 0.02$ & $0.00~\%$ & $0.00~\%$ & $0.00~\%$	 &   1.44 & 3.99 & 0.00 & 0.01 & +0.53 & $\bullet$ & $\bullet$ & $\bullet$ & \\
        $ k^*$	&  $0.71 \pm 0.07$ & $0.89 \pm 0.05$ & $1.00 \pm 0.04$ & $0.00~\%$ & $0.00~\%$ & $12.50~\%$	 &   1.93 & 2.81 & 0.01 & 0.01 & +0.26 & & $\bullet$ & $\bullet$ & \\
        $\gamma_s$	&  $0.43 \pm 0.14$ & $0.51 \pm 0.06$ & $0.47 \pm 0.07$ & $0.00~\%$ & $0.00~\%$ & $0.00~\%$ & 2.53 & 2.26 & 0.88 & 0.69  & -0.49 & $\bullet$ & $\bullet$ & $\bullet$ & $\bullet$ \\
        $ s $	    &  $1.32 \pm 0.18$ & $1.13 \pm 0.03$ & $1.07 \pm 0.02$ & $0.00~\%$ & $0.00~\%$ & $0.00~\%$	&   5.06 & 8.30 & 0.05 & 0.25 & -0.51 & $\bullet$ & $\bullet$ & $\bullet$ & \\
        $ s^*$	&  $2.09 \pm 0.84$ & $1.47 \pm 0.08$ & $1.33 \pm 0.10$ & $0.00~\%$ & $0.00~\%$ & $0.00~\%$	 &   7.18 & 5.60 & 0.48 & 0.62 & -0.39 & $\bullet$ &  $\bullet$ & $\bullet$ & $\bullet$ \\
        $m_{A}$	            & $0.09 \pm 0.04$ &  $0.12 \pm 0.04$ & $0.17 \pm 0.04$ & $0.00~\%$ & $0.00~\%$ & $0.00~\%$	 & 1.31 & 1.85 & 0.00 &0.00  & +0.02 & $\bullet$ & $\bullet$ & & \\
        $m_{B}$	            & $1.11 \pm 0.37$ & $1.54 \pm 0.11$ & $1.72 \pm 0.07$ & $0.00~\%$ & $0.00~\%$ & $0.00~\%$ 	 &   3.75 & 7.67 & 0.00 &0.00 & -0.09 & $\bullet$ &  $\bullet$ & $\bullet$ & \\
        $m_{C}$	            & $0.83 \pm 0.21$ & $1.19 \pm 0.10$ & $1.28 \pm 0.05$ & $18.75~\%$ & $0.00~\%$ & $0.00~\%$	&   2.30 & 6.04 & 0.00 &0.00 & +0.04 & & $\bullet$ & $\bullet$ & \\
        $m_{D}$	            & $0.22 \pm 0.09$ &  $0.27 \pm 0.11$ & $0.37 \pm 0.06$ & $0.00~\%$ & $0.00~\%$ & $0.00~\%$	&   0.97 & 2.45 & 0.00 &0.00 & +0.24 & $\bullet$ & $\bullet$ & &\\
        $m_{E}$	                & $0.76 \pm 0.18$ & $1.27 \pm 0.16$ & $1.03 \pm 0.06$ & $12.50~\%$ & $6.25~\%$ & $18.75~\%$ &   1.66 & 0.72 & 0.00 & 0.00 & -0.23 &  & $\bullet$ & & \\
        $m_{F}$	            & $0.24 \pm 0.07$ &  $0.37 \pm 0.05$ & $0.39 \pm 0.06$ & $0.00~\%$ & $0.00~\%$ & $0.00~\%$ 	&   1.87 & 1.80 & 0.00 &0.00 & -0.20 & $\bullet$ & $\bullet$ & &\\
        $m_{G}$	            & $0.36 \pm 0.14$ & $0.47 \pm 0.09$ & $0.56 \pm 0.05$ & $0.00~\%$ & $0.00~\%$ & $0.00~\%$	 &   1.82 & 4.43 & 0.00 &0.00 & +0.14 & $\bullet$ &  & & \\
        $m_{H}$	            &  $0.71 \pm 0.24$ & $1.25 \pm 0.11$ & $1.16 \pm 0.11$ & $0.00~\%$ & $0.00~\%$ & $0.00~\%$	&   2.67 & 3.66 & 0.00 &0.00 & -0.17 & $\bullet$ & $\bullet$ & $\bullet$ & \\
        $m_{I}$	            & $0.20 \pm 0.07$ &  $0.32 \pm 0.05$ & $0.36 \pm 0.05$ & $0.00~\%$ & $0.00~\%$ & $0.00~\%$	 & 1.68 & 2.48 & 0.00 &0.00 & -0.14 & $\bullet$  & $\bullet$ & & \\
        $m_{J}$        	    & $0.45 \pm 0.17$ &  $0.57 \pm 0.12$ & $0.73 \pm 0.05$ & $0.00~\%$ & $0.00~\%$ & $0.00~\%$	&   1.76 & 5.19 & 0.00 &0.00 & +0.11 & $\bullet$ & $\bullet$ & & \\
        $m_{K}$	            & $0.59 \pm 0.25$ &  $1.22 \pm 0.16$ & $1.02 \pm 0.08$ & $0.00~\%$ & $12.50~\%$ & $18.75~\%$ & 2.55 & 5.29 & 0.00 &0.00 & -0.24 & & $\bullet$ & $\bullet$ & \\
        $m_{L}$	            & $0.03 \pm 0.02$ & $0.04 \pm 0.02$ & $0.06 \pm 0.02$ & $0.00~\%$ & $0.00~\%$  & $0.00~\%$	& 1.53 & 1.85 & 0.04 &0.35 & +0.10 & $\bullet$ & $\bullet$ & & \\
        $m_{M}$               & $0.26 \pm 0.10$ & $0.39 \pm 0.06$ & $0.46 \pm 0.08$ & $0.00~\%$ & $0.00~\%$ & $0.00~\%$	 & 2.11 & 2.16 & 0.00 &0.00 & -0.14 & $\bullet$ & $\bullet$ & $\bullet$ & \\
        \hline
        \end{tabular}
\end{table*}


\subsection{Distinguishing books from shuffled sequences}

Our first aim is to identify measurements capable of distinguishing between natural and shuffled texts, which will be referred to as informative measurements. For instance, for $X=B$ in Fig.~\ref{img:example} all values are much smaller than 1 in all three sets of texts, indicating that this measurement takes smaller values in natural texts than in shuffled texts.
In order to quantify the distance of a set of values $\{X\}$ to $X=1$ we define the quantity $\rho(X=1,\{X\})$
as the proportion of elements in  the set $\{X\}$ for which $X=1$ lies within the interval $X \pm \epsilon(X)$, where $\epsilon(X)$ arises from fluctuations due to the randomness of the shuffling process as defined in eq.~(\ref{sssigma}).
This leads to condition $\zeta_1$:
\begin{itemize}
    \item [$\zeta_1$:] $X$ is said to be informative if $\rho(X=1,\{X\}) \rightarrow 0$ for $|\{X\}| \rightarrow \infty$,
\end{itemize}
where $\{X\}$ is a set of values $X$ obtained over different texts in different languages or texts, and $|\{X\}|$ is the number of elements in this set.

We now discuss the results obtained applying $\zeta_1$ (with $\rho(X=1,\{X\}) = 0$) for all three sets of texts in our database for each of
the measurements described in Sec.~\ref{definicao}. Measurements which satisfied $\zeta_1$ are indicated by a $\bullet$ in Tab.~\ref{app:tab.1}.
Several of the network measurements ($d$, $ L $, $ C $, $ k^*$ and motifs
$m_{C}$, $m_{E}$ and $m_{K}$) do not fully satisfy $\zeta_1$. Consequently they
cannot be used to distinguishing a manuscript from its shuffled version. This
finding is rather surprising because some of the latter measurements were proven
useful to grasp subtleties in text, e.g.\ for author
recognition~\cite{comparing}. In the latter application, however, the networks
representing text did not contain stopwords and the texts were lemmatized.
The averaging over the $50$ most frequent words seems to be essential to satisfy
$\zeta_1$  for the clustering coefficient and for the shortest paths (note that
$C^*$ and $L^*$ are informative while $C$ and $L$ are not). This means that the informativeness of these quantities is concentrated in the most frequent words. On the other hand, for the degree, an opposite effect occurs, i.e., $k$ is
informative and $ k^*$ is not. The informativeness of
intermittency ($I$ and $I^*$) may be explained by the fact 
by construction $I_i \simeq 1$ in shuffled texts (see Sec.~\ref{sec:intermittency}). Because in natural texts many words tend to appear clustered in regions $I_i > 1$ and $I_i^* > 1$.
The selectivity $s$ is also strongly affected by the shuffling process.
Words in shuffled texts tend to be less selective, which yields an increase in
$\gamma_s$~\cite{selet} (i.e., very selective words occur very sporadically) and
a decrease in $ s $ and $ s^*$. The selectivity is
related to the effect of word consistency~\footnote{Consistent words are those
tending to co-occur always with the same words, and therefore they are also
selective words.} (see Ref.~\cite{jstat}) which was verified to be common in
English, especially for very frequent words. The number of bigrams
$B$ is also informative, which means that in natural languages
it is unlikely that the same word is repeated (when compared with random texts).
As for the informative motifs, $m_{A}$, $m_{D}$, $m_{F}$, $m_{G}$, $m_{I}$, $m_{J}$, $m_{L}$ and $m_{M}$ rarely occur in natural language texts ($\langle X \rangle < 1$) while motif $m_{B} $ was the only measurement taking values
above and below $1$. The emergence of this motif therefore appears to depend on
the syntax, being very rare for Xhosa, Vietnamese, Swahili, Korean, Hebrew and
Arabic.

\subsection{Dependence on style and language}

We are now interested in investigating which text-measurements are more dependent on the language than on the style of the book, and
vice-versa. Measurements depending predominantly on the syntax are expected to have larger variability across languages than across texts. On
the other hand, measurements
 depending mainly on the story (semantics) being told are expected to have
 larger variability across texts in the same language, i.e.\ $t=\tau$~\footnote{This approach could be extended to account for different text genres, for distinct characteristics could be expected from novels, lyrics, encyclopedia, scientific texts, etc.}.
The variability of the measurements
 was computed with the coefficient of variation $\upsilon = \sigma(X) / \langle X \rangle$, where $\sigma(X)$
and $\langle X \rangle$ represent respectively the standard deviation and the average computed for the books in the set $\{X\}$. Thus, we may
assume that $X$ is more dependent on the language than on the style/semantics if condition $\zeta_2$ is satisfied:
\begin{itemize}
    \item [$\zeta_2$:] $X$ is more dependent on the language (or syntax) than it is on the style (or semantics) if $\upsilon_{t=\tau,l} > \upsilon_{t,l=\lambda}$.
\end{itemize}
Measurements failing to comply with condition $\zeta_2$ have $\upsilon_{t,l=\lambda} > \upsilon_{t=\tau,l}$ and therefore are more dependent on the
style/semantics than on the language/syntax. In order to quantify whether $\upsilon_{t=\tau,l} > \upsilon_{t,l=\lambda}$ or $\upsilon_{t,l=\lambda} > \upsilon_{t=\tau,l}$ is statistically significant, we took the confidence interval of $\upsilon_{t=\tau,l}$ and $\upsilon_{t,l=\lambda}$. Let $\iota(\upsilon)$
be the confidence interval for $\upsilon$ computed using the noncentral t-distribution~\cite{noncentral}, then $\zeta_2$ is valid if there is little intersection of the confidence intervals. In other words:
\begin{itemize}
    \item [$\zeta_2'$:] The inequality  $\upsilon_{t=\tau,l} > \upsilon_{t,l=\lambda}$ (or $\upsilon_{t,l=\lambda} > \upsilon_{t=\tau,l}$)  is
    valid only if $\iota(\upsilon_{t=\tau,l})~\cap~\iota(\upsilon_{t,l=\lambda}) \rightarrow 0$ for $|\{X\}| \rightarrow \infty$.
\end{itemize}
The confidence intervals were assumed to have little intersection if
$\iota(\upsilon_{t=\tau,l})~\cap~\iota(\upsilon_{t,l=\lambda}) \leq 0.05 \times
\iota(\upsilon_{t=\tau,l})~\cup~\iota(\upsilon_{t,l=\lambda})$. We took a
significance level $\alpha = 0.95$ in the construction of the confidence
intervals.

The results for the measurements satisfying conditions $\zeta_2$ and $\zeta_2'$ are shown in Tab.~\ref{app:tab.1}. Measurements satisfying conditions $\zeta_2$ and $\zeta_2'$ serve to examine the dependency on the syntax or on the style/semantics. The vocabulary size $M$,
and the network measurements $r$, $L$, $L^*$, $C$, $k$ and $k^*$ are more dependent on syntax than on semantics. The measurements derived from the selectivity ($\gamma_s$, $s$ and $s^*$) are also strongly dependent on the language. With regard to the motifs, five of them satisfy $\zeta_2$ and $\zeta_2'$: $m_{B}$, $m_{C}$, $m_{H}$, $m_{K}$ and $m_{M}$. Remarkably, $ I $ and $ I^*$ are the only measurements with low values of $\upsilon_{t=new,l} / \upsilon_{t, l=\lambda}$.
%
Reciprocally, the only measurement which statistically significantly violated $\zeta_2$ (i.e., satisfied $\zeta_2'$) was $I^*$. This confirms that the average intermittency of the most frequent words is more dependent on the style than on the language.

\subsection{On the representativeness of measurements}

{The practical implementation of our general framework was done quantifying the variation across languages using a single book (the New Testament). This was done because of the lack of available books in a large number of languages. In order for this approach to work it is essential to determine whether} fluctuations across different languages are representative of the fluctuations observed in different books. We now try to determine the measurements $X$ whose {\em actual values} of a single book on a specific language~$\lambda$ ($X_{t=new,l=\lambda}$)  are compatible to other books in the same language ($X_{t,l=\lambda}$). To this end we define the compatibility $c(X,P)$ of $X_{t=new,l=\lambda}$ to $P(X_{t,l=\lambda})$. The distribution $P$ was taken with the Parzen-windowing interpolation~\cite{parzening} using a Gaussian function as kernel. More precisely, $P$ was constructed adding Gaussian distributions centered around each $X$ observed over different texts in a fixed language~$\lambda$. Mathematically, the compatibility $c(X,P)$ is computed
as
\begin{equation} \label{compatibilidade}
c(X,P) = \left\{
\begin{array}{ll}
     2 \times \int_{-\infty}^{X^*} P(X) dX  & \text{if}~X < X_{\text{median}}, \\
     2 \times \int_{X^*}^{+\infty} P(X) dX  & \text{if}~X \ge X_{\text{median}}, \\
\end{array}
\right.
\end{equation}
where $X_{\text{median}}$ is the median of $P(X)$. For practical purposes, we consider that $X_{t=new,l=\lambda}$ is compatible with other books written in the same language $\lambda$ if $\zeta_3$ is fulfilled:
\begin{itemize}
\item [$\zeta_3$:] $X_{t=new,l}$ is a representative measurement of the language $\lambda$ if $c(X_{t=new,l=\lambda},P(X_{t,l=\lambda})) > 0.05$.
\end{itemize}
The representativeness of the measurements computed for the New Testament was
checked using the distribution $P(X)$ obtained from the set of books written in
Portuguese and English. The standard deviation employed in the Parzen method was
the worst deviation between English and Portuguese, i.e.\ $\sigma = \min \{ \sigma_{\text{pt}},\sigma_{\text{en}}\}$.
The measurements satisfying $\zeta_3$ for both English and Portuguese datasets are displayed in the last column of Tab. \ref{app:tab.1}. With regard to the network measurements, only $L$, $L^*$, $C$ and $C^*$ are representative, suggesting that they are weakly dependent on the variation of style (obviously assuming the New Testament as a reference).
In addition, $I$, $I^*$, $B$, $\gamma_s$, $s^*$ and $m_L$ turned out to be representative measurements.

\section{Case Study: the Voynich Manuscript (VMS)}  \label{results}

{So far we have introduced a framework for identifying the dependency of different measurements on the language and story of different books. We now investigate which extent the measurements we identified as relevant can provide information on analysis of single texts.} The Voynich Manuscript (VMS), named after the book dealer Wilfrid Voynich who bought the book in the early XX century, is a $240$ page folio that dates back to the XV century. Its mysterious aspect has captivated people's attention for centuries. Indeed, VMS has been studied by professional cryptographers, being a challenge to scholars and decoders~\cite{vms,hoaxref}, currently included among the six most important ciphers~\cite{vms}. The various hypotheses about VMS can be summarized into three categories:
(i) A sequence of words without a meaningful message; (ii) a  meaningful text written originally in an existing language which was coded (and possibly encrypted) in the Voynich alphabet; and (iii) a meaningful text written in an unknown (possibly constructed) language. While it is impossible to investigate systematically all these hypotheses, here we perform a number of statistical analysis which aim at clarifying the feasibility of each of these scenarios. To address point (i) we analyze shuffled texts. To address point (ii) we consider $15$ different languages, including the artificial language Esperanto that allows us to touch on point (iii) too. We do not consider the effect of encryption of the text.

The statistical properties of VMS were obtained to try and answer the questions posed in Tab.~\ref{tab.questions}, which required checking the measurements that would lead to statistically significant results. To check whether a given text is compatible with its shuffled version, $X$ computed in texts written in natural languages should always be far from $X = 1$, and therefore only informative measurements are able to answer question {Q$_1$}. To test whether a text is consistent with some natural language (question {Q$_2$}), the texts employed as basis for comparison (i.e., the New Testament) should be representative of the language. Accordingly, condition $\zeta_3$ must be satisfied when selecting suitable measurements to answer {Q$_2$}. $\zeta_2$ and $\zeta_2'$ must be satisfied for measurements suitable to answer {Q$_3$} because the variance in style within a language should be small, if one wishes to determine the most similar language. Otherwise, an outlier text in terms of style could be taken as belonging to another language. An analogous reasoning applies to selecting measurements to identify the closest style. Finally, note that answers for Q$_3$ and Q$_4$ depend on a comparison with the New Testament in our dataset. Hence, suitable measurements must fulfill condition $\zeta_3$ in order to ensure that the measurements computed for the New Testament are representative of the language.

\begin{table*}
\centering
\caption{\label{tab.questions}The conditions that must be fulfilled by the
measurements for answering each of the Questions posed. For
Q$_1$, $X$ should not be close to $X=1$ because $X \approx 1$ in shuffled texts.
In the case of Q$_3$, it is desirable that there is no intersection between the measurements computed for books belonging to different languages. Therefore $\zeta_2$ and $\zeta_2'$ should be fulfilled. To find the closest style, the measurement must be strongly dependent on style, i.e.\ only $\zeta_2'$ should be fulfilled. Finally, if a question involves a comparison of the unknown manuscript with the New Testament then it requires that the measurements employed are representative. Therefore, Q$_2$, Q$_3$ and Q$_4$ require the fulfillment of condition $\zeta_3$.}
\begin{tabular}{|c|c|c|c|c|c|}
\hline
\multicolumn{2}{|c|}{\bf Questions} & $\zeta_1$ & $\zeta_2$ & $\zeta_2'$ & $\zeta_3$  \\
\hline
{Q$_1$} &  Is the text compatible with shuffled version? & $\bullet$ & & &  \\
\hline
{Q$_2$} &  Is the text compatible with a natural language? & & & & $\bullet$ \\
\hline
{Q$_3$} &  Which language is closer to the manuscript? & & $\bullet$  &  $\bullet$  & $\bullet$  \\
\hline
{Q$_4$} &  Which style is closer to the manuscript? & & & $\bullet$  & $\bullet$  \\
\hline
\end{tabular}
\end{table*}


\subsection{Is the VMS distinguishable from its shuffled text?}

Before checking the compatibility of the VMS with shuffled texts, we verified if {Q$_1$} can be accurately answered in a set of books written in Portuguese and English, henceforth referred to as test dataset (see SI-Tab.~3). A given test text was considered as not shuffled if the interval $X - \epsilon(X)$ to $X + \epsilon(X)$ does not include $X=1$. To quantify the distance of a text from its shuffled version, we defined the distance $D$:
\begin{equation} \label{eq:dist}
D = \frac{|X - 1|}{\epsilon(X)},
\end{equation}
which quantifies how many $\epsilon$'s the value $X$ is far from $X=1$. As one should expect, the values of $X$ computed in the test dataset for $\lambda = \text{pt}$ and $\lambda = \text{en}$ (see SI-Tab.~4) indicate that all texts are not compatible with its shuffled version because $D > 1$, which means that the interval from $X - \epsilon(X)$ to $X + \epsilon(X)$ does not include $X=1$. Once the methodology appropriately classified the texts in the test dataset as incompatible with their shuffled versions, we are now in position to apply it to the VMS.

The values of $X$ for the VMS, denoted as $X_{\rm VMS}$, in Tab.~\ref{tab.00} indicate that the VMS is not compatible with shuffled texts, because the interval from $X_{\rm VMS} - \epsilon(X_{\rm VMS})$ to $X_{\rm VMS} + \epsilon(X_{\rm VMS})$ does not include $X=1$. All but one measurement ($C^*$) include $X=1$ in the interval $X_{\rm VMS} \pm \epsilon(X_{\rm VMS})$, suggesting that the word order in the VMS is not established by chance. The property of the VMS that is most distinguishable from shuffled texts was determined quantitatively using the distance $D_{\rm VMS}$ from eq. (\ref{eq:dist}).
Tab.~\ref{tab.00} shows the largest distances for intermittency ($I$ and $I^*$) and network measurements ($k$ and $L^*$). Because intermittency is strongly affected by stylistic/semantic aspects and network measurements are mainly influenced by syntactic factors, we take these results to mean that the VMS is not compatible with shuffled, meaningless texts.

\begin{table*}
    \centering
    \caption{\label{tab.00}Values of $X$ for the Voynich Manuscript considering only the informative measurements (i.e., the measurements satisfying $\zeta_1$). Apart from $ C^*$ all measurements point to the VMS being different from shuffled texts.}
        \begin{tabular}{|c|c|c|c|c|c|c|c|c|c|c|c|c|c|c|c|c|c|c|c|c|}
        \hline
        $X$ & $ L^*$ & $ C^*$ & $ I $ & $ I^*$ & $B$ & $ k $ &  $\gamma_s$ &  $m_{G}$ & $m_{F}$ &  $m_{J}$ & $m_{D}$ & $m_{I}$  &  $m_{M}$  & $m_{A}$ & $m_{L}$ \\
        \hline
        $X_{\rm VMS} - \epsilon(X_{\rm VMS})$ & 1.069 & 0.981 & 1.423 & 1.875 & 2.333 & 0.948 & 0.617 & 0.782 & 0.738 & 0.784 & 0.908 & 0.724 & 0.783 & 0.728 & 0.549 \\
        $X_{\rm VMS}$ & 1.071 & 0.999 & 1.433 & 1.890 & 2.637 & 0.949 & 0.692 & 0.796 & 0.751 & 0.798 & 0.940 & 0.733 & 0.801 & 0.739 & 0.582 \\
        $X_{\rm VMS} + \epsilon(X_{\rm VMS})$ & 1.072 & 1.017 & 1.443 & 1.904 & 2.940 & 0.950 & 0.768 & 0.809 & 0.765 & 0.813 & 0.971 & 0.741 & 0.819 & 0.751 & 0.616 \\
        \hline
        $D_{\rm VMS}$ & 47 & 0 & 44 & 61 & 5 & 51 & 23 & 15 & 18 & 14 & 2 & 32 & 11 & 23 & 12 \\
        \hline
        \end{tabular}
\end{table*}

\subsection{Is the VMS compatible with a text in natural languages ?}

The compatibility with natural languages was checked by comparing the suitable measurements for the VMS with those for the New Testament written in $15$ languages. Similarly to analysis of compatibility with shuffled texts, we validated our strategy in the test dataset as follows. The compatibility with natural texts was computed using eq. (\ref{compatibilidade}), where $P$ was computed from the New Testament dataset. The standard deviation on each Gaussian representing a book in the test dataset should be proportionally to the variation of $X$ across different texts and therefore we used the worst $\sigma$ between English and Portuguese. {The values displayed in SI-Tab.~5} reveal that all books are compatible with natural texts, as one should expect. Therefore we have good indications the proposed strategy is able to properly decide whether a text is compatible with natural languages.

%

\begin{figure*}
\begin{center}
\includegraphics[width=0.9\textwidth]{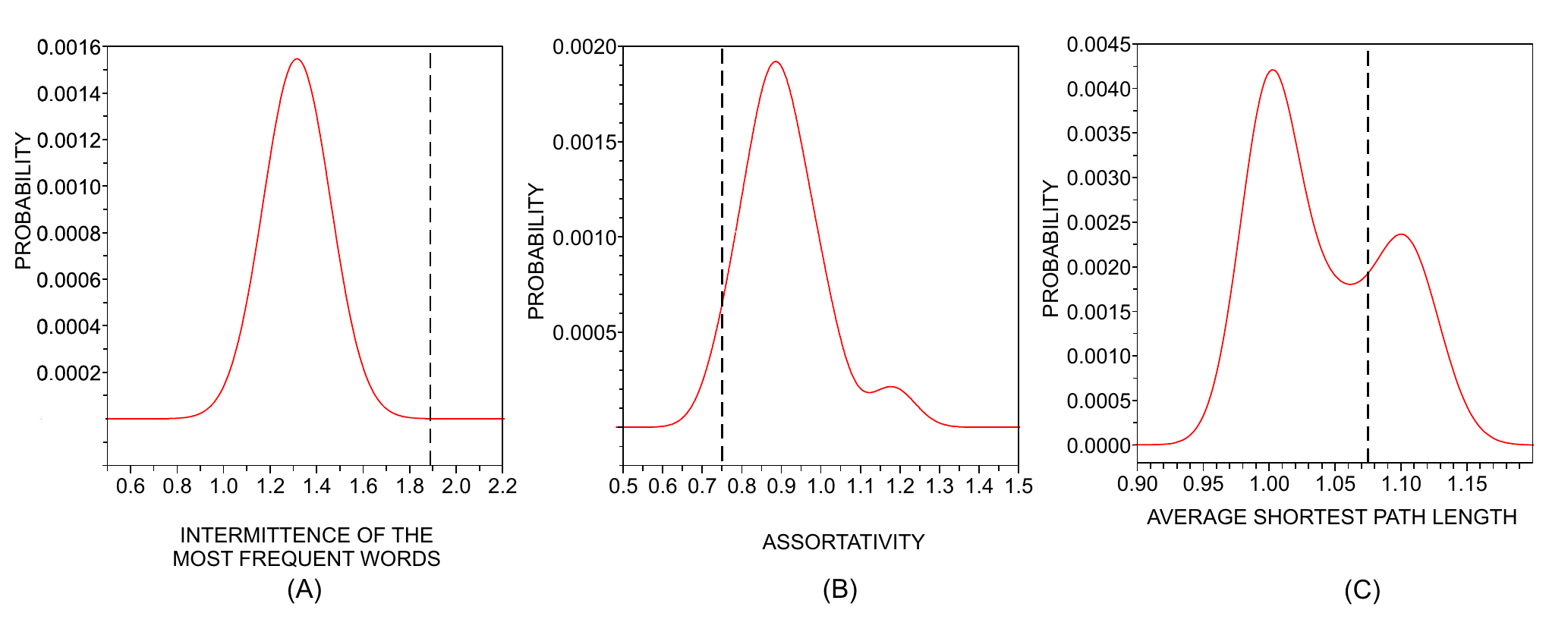}
\end{center}
\caption{\label{fig:network}
Distribution of measurements for the New Testament compared with the measurement
obtained for VMS (dotted   line). The measurements are
(a) $X = I^*$ (intermittency of the most frequent words);
(b) $X = r$ (assortativity) and
(c) $X = L$ (average shortest path length).
While in (a) VMS is not compatible with natural languages,
in (b) and (c) the compatibility was verified since $c(X_{\rm VMS},P) > 0.05$.
}
\end{figure*}

The distance from the VMS to the natural languages was estimated by obtaining the compatibility $c(X_{\rm VMS},P(X_{t=\text{new},l}))$ (see eq. \ref{compatibilidade}). In this case, $P$ was constructed adding Gaussian distributions centered around each $X$ observed in the New Testament over different languages~$\lambda$.
The distribution $P$ for three measurements is illustrated in Fig.~\ref{fig:network}. The values of $c(X_{\rm VMS},P(X_{t=\text{new},l}))$
displayed in Tab.~\ref{tab.vmscomp} confirm that VMS is compatible with natural languages for most of the measurements suitable to answer Q$_2$. The exceptions were $B$ and $I^*$. A large $B$ is a particular feature of VMS because the number of duplicated bigrams is much greater than the expected by chance, unlike natural languages. $I^*$ is higher for VMS than the typically observed in natural languages (see Fig.~\ref{fig:network}(a)), even though the absolute intermittence value of the most frequent words in VMS is not far from those for natural languages. {Since the intermittency $I$ is related to large scale distribution of a (key) word in the text, we speculate that the reason for these observations may be the fact that the VMS is a compendium of different topics.}
%

\begin{table}
    \centering
    \caption{\label{tab.vmscomp}Compatibility of VMS with natural languages. Except for $I^*$ and $B$, the measurements computed for VMS are consistent with those expected for texts written in natural languages.
    }
        \begin{tabular}{|c|c|c|c|c|c|c|c|c|c|c|}
        \hline
        {\bf $X$} & $r$ & $L$ & $L^*$ & $C$ & $C^*$ & $I$ & $I^*$ & $B$ & $s^*$ & $\gamma_s$ \\
        \hline
        $c$ & 0.14 & 0.62 & 0.99 & 0.96 & 0.05 & 0.39 & 0.00 & 0.00 & 0.09 & 0.12 \\
        \hline
        \end{tabular}
\end{table}

\subsection{Which language/style is closer to the VMS?}

We address this question in full generality but we shall show that with the limited dataset employed, we cannot obtain a faithful
  prediction of the language of a manuscript.  Given a text $\tau$, we identify the most similar language according to the following
  procedure. We first calculate the Euclidean distance (using the z-normalized values of the
  measurements suitable to answer Q$_3$ in Tab.~\ref{tab.questions}) between the book under analysis and the versions of the New
  Testament.  Let $R_{\lambda,\tau}$ be the ranking
  obtained by language $\lambda$ in the text~$\tau$. Given a set of texts $\mathcal{T}$ written in the same language, this procedure yields a list
  of $R_{\lambda,\tau}$ for each $\tau \in \mathcal{T}$. In this case, it is useful to combine the different $R_{\lambda,\tau}$ by
  considering the product of the normalized ranks
\begin{equation}\label{eq.delta}
    \delta_{\lambda} =  \prod_{\tau \in \mathcal{T}}\frac{R_{\lambda,\tau}}{|\mathcal{T}|},
\end{equation}
where $|\mathcal{T}|$ is the number of texts in the database $\mathcal{T}$. This choice is motivated by the fact that
$R_{\lambda,\tau}/|\mathcal{T}|$ corresponds to the probability of achieving by chance a rank as good as $R_{\lambda,\tau}$ so that
  $\delta_{\lambda}$ in Eq.~(\ref{eq.delta}) corresponds to the probability of obtaining such a ranking by chance {\em in every single}
  case. By ranking the languages according to $\delta_{\lambda}$ we obtain a rank of best candidates for the language of the texts in
  $\mathcal{T}$.

In our control experiments with $|\mathcal{T}|=15$ known texts we verified that the measurements suitable to answer Q$_3$ led to
  results for the books in
  Portuguese and English of our dataset which not always coincide with the correct language. In the case of the Portuguese test dataset,
  Portuguese was the second best language (after Greek), while in the English dataset the most similar languages were Greek and Russian and
  English was only in place $6$. Even though the most similar language did not match the language of the books, the $\delta_{\lambda}$
  obtained were significantly better than chance  (p-value=$1.0\;10^-7$ and $4.3\; 10^{-5}$,
  respectively in the English and Portuguese test sets).

The reason why the procedure above was unable to predict the accurate language is directly related to the use of only one example (a version
of the New Testament) for each language, while in robust classification methods many examples are used for each class. Hence, finding the
most similar language to VMS will require further efforts, with the analysis of as many as possible books representing each language, which
will be a challenge since there are not many texts widely translated into many languages.

\subsection{Keywords of the VMS}

One key problem in information sciences is the detection of important words as they offer clues about the text content. In the context of decryption, the identification of keywords may be helpful for guiding the deciphering process, because cryptographers could focus their attention on the most relevant words. Traditional techniques are based on the analysis of frequency, such as the widely used term frequency-inverse document frequency~\cite{manning} (tf-idf). Basically, it assigns a high relevance to a word if it is frequent in the document under analysis but not in other documents of the collection. The main drawback associated with this approach is the requirement of
a set of representative documents in the same language. Obviously, this restriction makes it impossible to apply tf-idf to the VMS, since there is only one document written in this ``language''. Another possibility would be to use entropy-based methods~\cite{entropy-keyword,kkk} to detect keywords. {However, the application of all these methods to cases such as the VMS will be   limited because they typically} require the manuscript to be arranged in partitions, such as chapters and sections, {which are not easily identified in the VMS}.

To overcome this problem, we use the fact that keywords show high intermittency inside a single text~\cite{key1,kkk}. {Therefore, this feature can play the role traditionally played by the inverse document frequency (idf). In agreement with the spirit of the tf-idf analysis}, we define the relevance $\Omega_i$ of word $i$ as proportional to both the intermittency and frequency as follows:
\begin{equation} \label{segunda}
    \Omega_i = (I_i - 1) \sqrt{\log N_i}.
\end{equation}
Note that with the factor $I_i$, words with $I \simeq 1$ receive low values of $\Omega$ even if they are very frequent.
There are other methods for detecting keywords relying on the analysis of the uneven distribution of the words~\cite{improving}, but we
decided not to use them because they generate better results for short texts, which is not the case of VMS.
{For the case of small texts and small frequency, corrections on our definition of intermittency should be used, see
Ref.~\cite{improving} which also contains alternatives methods for the computation of key-words from intermittency.}
In order to validate $\Omega$ we
applied Eq.~(\ref{segunda}) to the New Testament in Portuguese, English and
German. An inspection of Tab.~\ref{tab.0000} for Portuguese, English and German
indicates that representative words have been captured, such as the characters
``Pilates'', ``Herod'', ``Isabel'' and ``Maria'' and important concepts of the
biblical background such as ``nasceu'' (was born), ``céus''/``himmelreich''
(heavens), ``heuchler'' (hypocrite), ``demons'' and ``sabbath''.
{In the right column of Tab.~\ref{tab.0000} we present the list of `words obtained for the VMS through the same procedure, which are natural candidates as keywords. }

\begin{figure*}[bt]
  \centering
  \includegraphics[width=1.8\columnwidth]{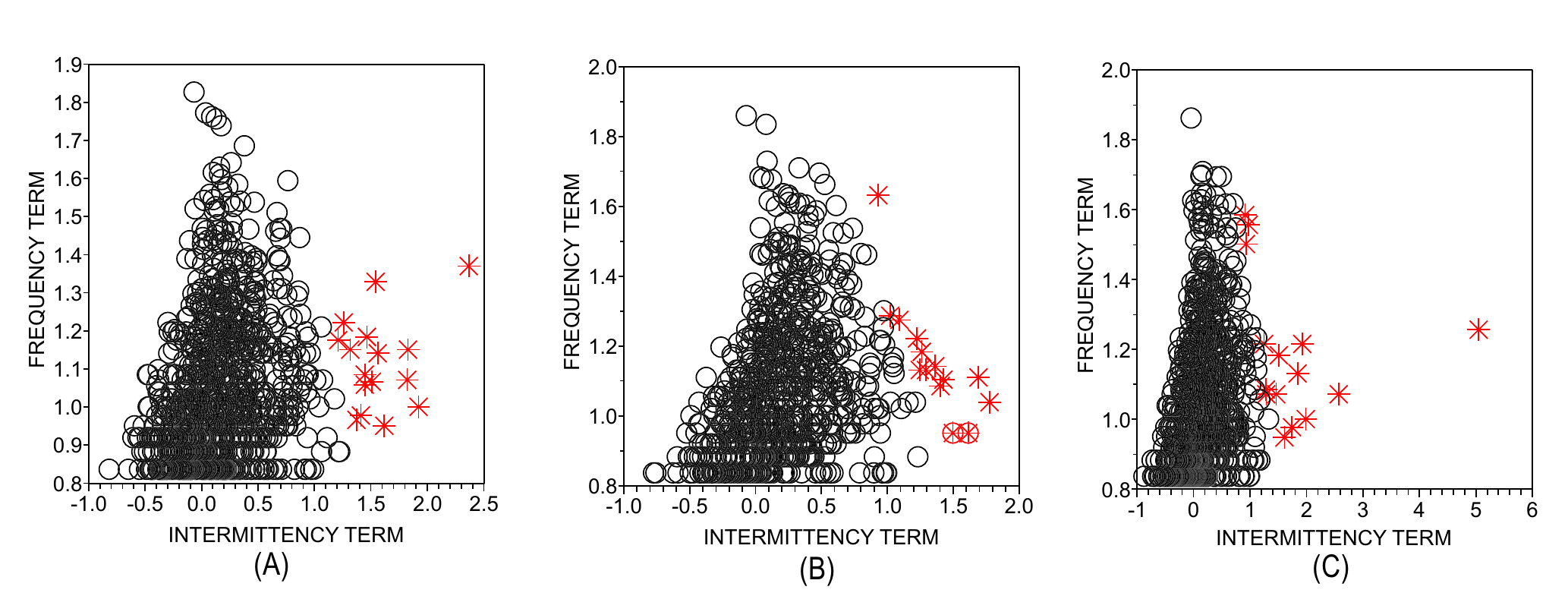}
  \caption{
  List of keywords (marked as $\ast$) found for the New Testament in (a) Portuguese; (b) English
and (c) German. The intermittency term refers to $(I_i - 1)$ and the frequency
term refers to $\sqrt{\log N_i}$ in eq. (\ref{segunda}). Note that keywords are
characterized by high intermittency and frequency terms.
  }
\end{figure*}

\begin{table}
    \centering
    \caption{\label{tab.0000} Keywords of the New Testament (English, Portuguese and German) and the VMS using Eq.~(\ref{segunda}).}
        \begin{tabular}{|c|c|c|c|}
        \hline
        Portuguese & English & German & Voynich \\
        \hline
        nasceu & begat & zeugete & cthy \\
        Pilatos & Pilates & zentner & qokeedy\\
        c\'eus &  talents & himmelreich & shedy\\
        bem-aventurados & loaves & pilatus & qokain\\
        Isabel & Herod & schwert & chor\\
        anjo & tares & Maria & lkaiin\\
        menino & vineyard & Elisabeth & qol\\
        vinha &  shall & Etliches & lchedy\\
        sumo & boat & unkraut & sho\\
        sepulcro & demons & euch & qokaiin\\
        joio & five & schiff & olkeedy\\
        Maria & pay & ihn & qokal\\
        portanto & sabbath & weden & qotain\\
        Herodes &  hear & heuchler & dchor\\
        talentos & whosoever & tempel & otedy\\
        \hline
    \end{tabular}
\end{table}

\section{Conclusion} \label{conclusion}

In this paper we have {developed the first steps towards a statistical} framework to determine whether an
unknown piece of text, recognized as such by the presence of a sequence of symbols organized in ``words'', is a meaningful text {and which language or style is closer to it}. The framework encompassed statistical analysis of individual words and then books using three types of measurements, namely metrics obtained from first-order statistics, metrics from networks representing text and the intermittency properties of words in a text. We identify a set of measurements capable of distinguishing between real texts and their shuffled versions, which were referred to as informative measurements. With further comparative studies involving the same text (New Testament) in 15 languages and distinct books in English and Portuguese, we could also find metrics that depend on the language (syntax) to a larger extent than on the story being told (semantics). Therefore, these measurements might be employed in language-dependent applications. Significantly, the analysis was based entirely on statistical properties of words, and did not require any knowledge about the meaning of the words or even
the alphabet in which texts were encoded.

The use of the framework was exemplified with the analysis of the Voynich Manuscript, with the final conclusion that it differs from a random sequence of words, being compatible with natural languages. Even though our approach is not aimed at deciphering Voynich, it was capable of providing keywords that could be helpful for decipherers in the future.

\begin{acknowledgments}

The authors are grateful to CNPq and FAPESP (2010/00927-9 and 2011/50761-2) for the financial support. DRA acknowledges support from the MPIPKS during his one-month visit to Dresden (Germany).

\end{acknowledgments}
%

\bibliography{template}

\end{document}